# An additive-manufactured microwave cavity for a compact cold-atom clock

Etienne Batori[1], Alan Bregazzi[2], Ben Lewis[2], Paul Griffin[2], Erling Riis[2], Gaetano Mileti[1,*] and Christoph Affolderbach[1]

[1] Université de Neuchâtel, Institut de Physique, Laboratoire Temps-Fréquence, Avenue de Bellevaux 51, 2000 Neuchâtel, Switzerland
[2] SUPA and Department of Physics, University of Strathclyde, G4 0NG, Glasgow, United Kingdom
* Corresponding author: gaetano.mileti@unine.ch

We present an additive-manufactured microwave cavity for a Ramsey-type, double resonance, compact cold-atom clock. Atoms can be laser cooled inside the cavity using a grating magneto-optic trap (GMOT) with the cavity providing an excellent $TE_{011}$-like mode while maintaining sufficient optical access for atomic detection. The cavity features a low Q-factor of 360 which conveniently reduces the cavity-pulling of the future clock. Despite the potential porosity of the additive-manufacturing process, we demonstrate that the cavity is well-suited for vacuum. A preliminary clock setup using cold atoms allows for measuring the Zeeman spectrum and Rabi oscillations in the cavity which enables us to infer excellent field uniformity and homogeneity respectively, across the volume accessed by the cold atoms. Ramsey spectroscopy is demonstrated, indicating the cavity is suitable for clock applications. Finally, we discuss the limitations of the future clock.

## I. INTRODUCTION

In the field of atomic clocks and frequency standards, the application of laser cooling and trapping has brought significant breakthroughs, in particular with cold atomic fountain clocks[1]. It is also widely used in more portable or compact industrial clocks[2,3] and gravimeters[4,5]. In the meantime, Ramsey-type optical-microwave double resonance (DR) clocks based on hot vapor-cells have been studied extensively for future portable[6], compact and space-grade[7,8] and miniature cell[9] clocks. Although they achieve state-of-the-art stabilities, they remain limited by a variety of effects such as strong coupling with fluctuating external parameters[6,10], vapor-cell frequency aging[11] and high relaxation rates. Similar limitations apply for vapor-cell coherent population trapping (CPT) clocks. Cold-atom CPT and DR frequency standards are hence excellent candidates for the next generation of compact atomic clocks.

To reduce the cooling and trapping complexity for reduced size, weight and power (SWaP) applications, grating magneto-optical traps (GMOTs) can be used to create a 3D MOT[12]. In this case, a silicon grating is etched to diffract one incoming beam into three opposing beams. The four beams allow the creation of a 3D MOT. This technology has already been applied to CPT Ramsey clocks[13–15] and a CPT mini-fountain[16]. However, CPT clocks low contrast and strong light-shift sensitivities make them less stable in the short and long-term, respectively. Unlike in miniature cell clocks, portable and high-performance CPT and DR Ramsey clocks based on vapor-cell or cold atoms share similar SWaPs, which eventually favors the latter at the price of developing a microwave cavity allowing for cooling and trapping the atoms.

To date, different microwave cavity geometries for compact cold-atom and vapor-cell clocks have been used: spherical[2], cylindrical[17,18] and loop-gap resonator (LGR)[19–22]. LGR techniques allow the creation of sub-wavelength, and hence more compact, cavities as small as a few mm wide[23]. Additive manufacturing (AM) of such microwave cavities for compact vapor-cell clocks has already been demonstrated[24,25] with typical quality factor of $Q_c \approx 70$. While a high cavity quality factor favors low phase shift and hence more accurate clocks[19], a low quality factor lowers the cavity pulling effect[26] which is of great interest for realizing more stable clocks.

Here, we present the conception, simulation and realization of a vacuum-grade aluminum AM microwave cavity for a compact $^{87}$Rb cold-atom clock. The atom cooling and trapping are achieved using only one laser beam incident on a GMOT hosted inside the cavity.

## II. CLOCK CONCEPT AND CAVITY

### a) Concept

In DR Ramsey $^{87}$Rb clocks, the clock transition with $\nu_0 = 6.835$ GHz (See Figure 1c) is interrogated with light and microwaves in a pulsed manner. After a trapping and cooling stage for a time $t_{load}$ a short optical state selection stage is utilized to prepare the atoms in the relevant initial clock state for interrogation. This state preparation beam propagates perpendicular to the cavity axis. The atoms are then driven between the atomic levels $|F = 2, m_F = 0\rangle$ and $|F = 1, m_F = 0\rangle$ using two $\pi/2$ microwave pulses of length $t_m$ separated by the Ramsey time $T_R$. The atomic populations are subsequently read out using an absorption method, with an additional beam, also aligned perpendicular to the cavity axis co-parallel to the state selection beam, during a time $t_{det}$. This concept is illustrated in Figure 1a and b.

In these conditions, the so-called light-shift effects on the clock transition are minimized compared to CW or CPT operation where the light is on during the clock part of the cycle. In principle, the short-term stability of the clock linearly improves with $t_m$ and $T_R$ as the atomic transition quality factor $Q_a$ equals[27]:

$$Q_a = 2\nu_0 \left( T_R + \frac{4}{\pi} t_m \right) \quad (1)$$

and the short-term stability of a signal-to-noise ratio (SNR)-limited clock is estimated as:

$$\sigma(\tau) = \frac{1}{\pi C Q_a} \frac{1}{SNR} \sqrt{\frac{T_C}{\tau}}, \quad (2)$$

with $T_C = t_{load} + t_{prep} + 2t_m + T_R + t_{det}$ the total cycle time and $C$ the fringe contrast.

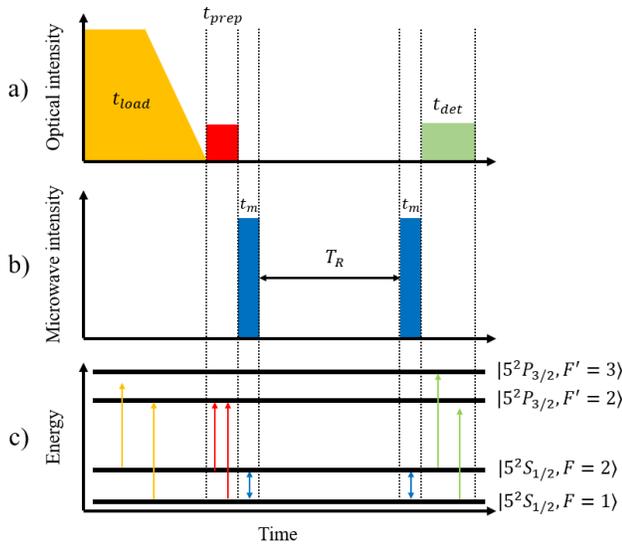

Figure 1 – a) Optical intensity with respect to time. Pulses lengths are not to scale. The different pulses heights represent the differences in intensities. b) π/2 microwave pulses separated by Ramsey time $T_R$. c) Relevant ⁸⁷Rb D2 line energy levels with corresponding excitation frequencies with respect to time.

Cold atoms allow for increased interrogation times $T_R$ due to low atomic speed (low atom temperature) and hence extended interrogation time in a finite interaction region. High-performance vapor-cell Ramsey operations typically show Ramsey times of the order of 3 ms[6] whereas a GMOT CPT mini-fountain demonstrated $T_R$ as long as 100 ms[16].
Atom trapping and cooling is achieved using a diffraction grating, placed at the bottom of the cavity (see Figure 2). In this geometry, a single input beam is sufficient. Two side holes in the cavity body are placed at the relevant height to prepare and probe the atom cloud. This simplifies the cavity design requirements as only three holes instead of the traditional six are necessary. Furthermore, the flat grating allows to interrogate the atoms right after the MOT sequence which would not be possible in a pyramid structure[20]. In principle, the atoms can be interrogated either with an absorption or fluorescence measurement.

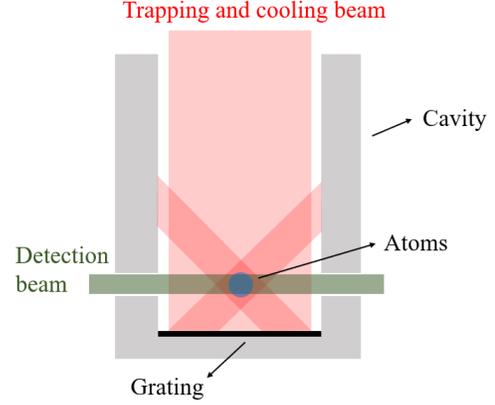

Figure 2 – Concept of a microwave cavity with integrated grating. The ensemble is placed in a vacuum chamber.

b) **Cavity design**

The LGR cavity geometry model is presented in Figure 3. With this electrode structure, cavities much smaller than the wavelength of the mode can be realized which is of great interest for compact physics packages. The use of additive-manufacturing allows for greater control and repeatability on such exotic electrode structures which cannot be easily attained with traditional manufacturing, especially when it comes to producing unique and small batches. A coarse estimate of the cavity resonance frequency $\nu_c$ can be made using eq. (3) with $n$ the number of electrodes, $R$ and $r$ the outer and inner diameters and $t$ and $w$ the width and thickness of the gaps[28]:

$$\nu_c = \frac{1}{2\pi} \sqrt{\frac{nt}{\pi r^2 \epsilon \mu w}} \sqrt{1 + \frac{r^2}{R^2 - (r+w)^2}} \sqrt{\frac{1}{1 + 2.5\frac{t}{w}}}. \quad (3)$$

With $\epsilon$ [F/m] and $\mu$ [N/A²] the permittivity and permeability inside of the cavity. Due to the absence of a vapor cell, the design is slightly simpler and the values for vacuum permittivity and permeability are used. Although eq. (3) yields the correct order of magnitude for the mode frequency, it is necessary to take the full geometry into account such as the height of the cavity $h$, the electrode support angle $\alpha$ (See Figure 1), the cavity material and the placement and geometry of the microwave coupling loop.
The grating is excluded from the simulations for the sake of time saving and because it has shown little impact on the final mode frequency. The cavity geometry is slightly different from previous vapor-cell designs with the addition of newer constraints. First, the presence of two 4-mm diameter side holes for the purpose of MOT diagnostics, state preparation and detection at the position of the atom cloud. Second, an inner cavity diameter that is sufficiently large to house the 2 cm x 2 cm grating chip and an upper aperture wide enough to allow the alignment of the trapping and cooling beam onto the grating surface.

These last requirements mainly put lower limit constraints on $R$ and $r$.

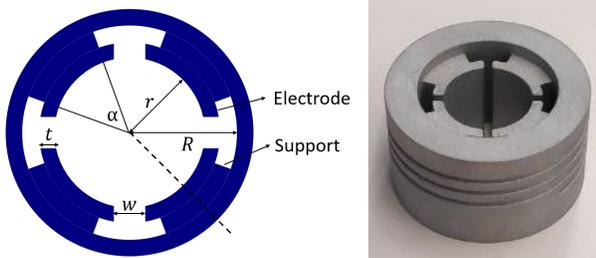

*Figure 3 – LEFT: Planar view of the LGR geometry. Dashed line indicates the zero-degree reference for the microwave coupling loop, see Figure 8. RIGHT: AM cavity body prior to holes being drilled.*

Eigenfrequency analysis sweeping $n$, $h$, $\alpha$, $R$, $r$, $w$ and $t$ is first performed to find a suitable set of parameters yielding a $TE_{011}$-like mode. The distance to the closest neighboring mode should be much larger than the full width at half maximum (FWHM) of the mode. An example of such a sweep is shown in Figure 4, sweeping $t$ and $w$ which shows that a large set of $(t, w)$ combinations is suitable for realizing the cavity indicated by the area between the two white contours. The figure shows the chosen combination which favors high $t$ and $w$ so that eventual manufacturing inaccuracies have smaller impact. For all suitable configurations, the closest neighboring mode is at least 800 MHz away.

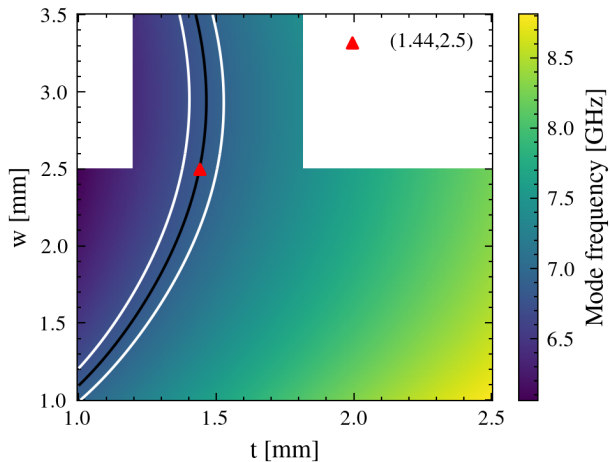

*Figure 4 - Best $TE_{011}$-like mode frequency [GHz]. The black line represents configurations with 6.8347 GHz mode frequency. The white lines are ± 100 MHz boundaries that represent the range that can be recovered by adjusting the cavity height.*

The expected magnetic field $H_z$ component and phase along the central axis of the cavity above the grating are presented in Figure 5. In the current experimental design, the atoms are prepared above the grating and then fall towards it in the least uniform part of the cavity field. Future designs should make use of the most uniform part of the cavity by pushing the atoms away from the grating. In all cases, as the field spatially varies between the two points where the atoms are interrogated, the pulses might have to show different injected microwave power or duration to ensure $\pi/2$ pulses [19]. In a sphere of 1 mm diameter – which is of the order of the initial size of the atomic cloud - at its initial position, the $H_z$ field and phase show deviations of 4% and 42 ppm, respectively.

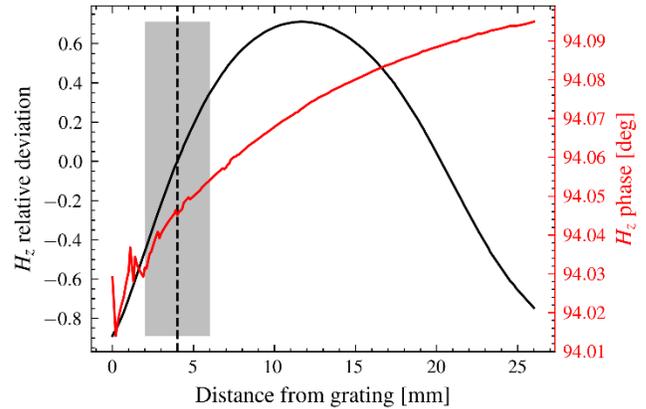

*Figure 5 – $H_z$ amplitude deviation and phase above the grating. The dashed line represents the initial atom cloud position. The deviation is normalized by the value of the $H_z$ field at the initial position of the atoms. The gray area shows the region accessible to the first version of the clock setup.*

The final cavity model is presented on Figure 6, to scale. The cavity body shown on Figure 3 was manufactured by a commercial additive manufacturing service, using standard CL31AL material according to DIN EN 1706 AlSi10Mg(b) while the other simpler and less critical parts were manufactured in our workshop in 6082 Al alloy.

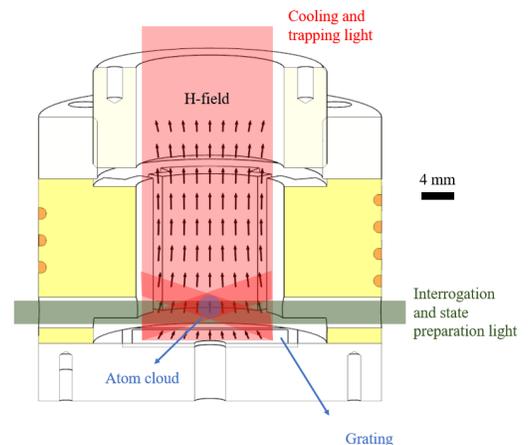

*Figure 6 – Final cavity model, to scale.*

### c) Mode characterization

Figure 7 compares the simulated and measured cavity $S_{11}$ microwave reflection spectrum over a wide frequency range. The simulation and measurement agree well on the mode positions with a small discrepancy of 60 MHz for the mode of interest which can be easily compensated by adjusting the height of the cavity. As predicted by the simulation, the left and right neighboring modes are 1.2 and 1.8 GHz away, respectively. They do not overlap with the mode of interest so that the field geometry is not degraded. The 8.5 GHz resonance is associated with a mode whose field energy is concentrated around the microwave coupling loop and low sensitivity to the other cavity parameters. The loop is realized as a small, rectangular copper wire loop of approximately 9 mm² cross-section with one end connected to the SMA core[29]. According to our simulations, the precise loop geometry does not strongly influence the mode of interest but rather the 8.5 GHz mode which is of no major concern for our application due the large distance between the two modes. However, the loop position relative to the electrodes is quite critical for efficient coupling of microwave radiation to the cavity, as illustrated by Figure 8.

*Table 1 – Cavity mode FWHM and Q-factor simulation vs measurement comparison.*

|  | FWHM [MHz] | $Q_c$ | Depth [dB] |
|---|---|---|---|
| Simulation | 4 | 1709 | -23 |
| Measurement | 19 | 360 | -9 |

The rather low quality factor minimizes the cavity pulling effect, indeed, the cavity-pulling shift [6,26]:

$$\Delta \nu_{cp} = -\frac{4}{\pi} \frac{Q_c}{Q_a} c(\theta) \Delta \nu_c, \quad (4)$$

with $Q_c$ and $Q_a$ the cavity and atomic transition quality factors, respectively, $c(\theta)$ a function of the microwave pulse area that can be made < 0.005[26] and $\Delta \nu_c$ the cavity detuning which, under vacuum, only depends on the cavity temperature. The temperature coefficient is measured as $\Delta \nu_c / \Delta T_{cav} = (-156 \pm 9)$ kHz/K which is similar to other models of AM cavities [6,25].

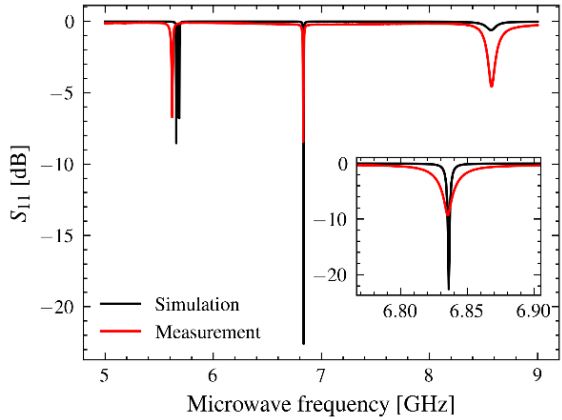

*Figure 7 – Simulated and measured $S_{11}$ spectrum of the cavity. The main figure and the inset share the same axis units.*

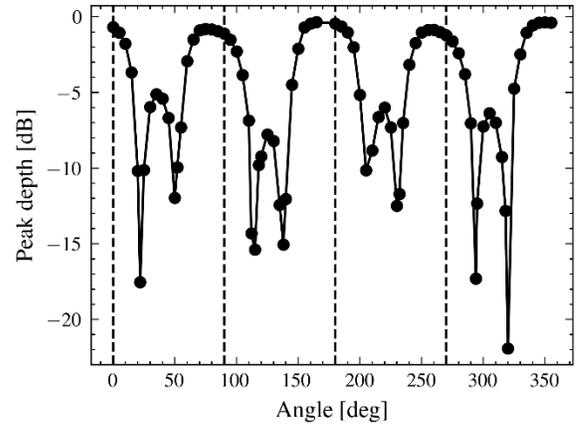

*Figure 8 – Measured microwave mode peak depth with respect to the rotation angle between the microwave coupling loop and the cavity. See Figure 3 for the 0° reference position. Given the angle, coupling to the mode of interest can be either enhanced or depleted.*

The simulated and measured $S_{11}$ of the 6.835 GHz mode (see inset of Figure 7 and Table 1) differ in depth and FWHM. This is explained first because the simulation did not include any wall roughness which is not realistic given the AM process. Second, for the sake of simulation time the grating was not included in the simulation. With its etched nature, it exhibits increased impedance to surface currents which also participates to power losses and hence line broadening. Lastly, the loss of mode amplitude is attributed to the high sensitivity of the mode to the angle between the microwave coupling loop and the electrodes, see the measured data in Figure 8. Although the angle was manually optimized to get a peak depth around -17 dB, corresponding to the loop placed approximately under one of the electrode wings, it was not perfectly preserved when securing the cavity, as evidenced by Figure 7.

## III. SETUP

### a) Experimental clock apparatus

The cavity is mounted in a commercial vacuum chamber, with the grating placed at the bottom of the cavity. Resistively heated dispensers provide the necessary $^{87}$Rb vapor. The outgassing properties of the aluminum-based additively-manufactured material were studied in a dedicated test chamber using the gas accumulation method[30]. Its outgassing properties were compared to a sample of bulk aluminum. Both samples show similar outgassing rates estimated around $7 \times 10^{-9}$ mbar m³ s⁻¹ m⁻² which suggests little to no degradation caused by the additive-manufacturing material compared to traditional aluminum alloy.

Furthermore, measurement of the cold-atom cloud 1/e lifetime[31] gives an estimate of the pressure of the order of $1 \times 10^{-8}$ mbar which also indicates this material suitability for in-vacuum cold-atom applications.

The magnetic field is controlled with 3 orthogonal sets of Helmholtz coils placed outside of the chamber to minimize the field in the grating plane and apply a C-field along the principal axis of the cavity to separate the Zeeman sublevels. A pair of anti-Helmholtz coils are placed inside the vacuum chamber allow trapping and cooling. The cavity loop is connected to a commercial microwave generator (Keysight E8257D).

A single 780 nm extended cavity diode laser (ECDL), frequency stabilized with saturated absorption spectroscopy is used throughout. This light is passed through an electro-optical modulator (EOM) to allow the generation of optical sidebands on the carrier frequency for hyperfine repumping of the atoms. A tapered amplifier is then used to amplify the light before dividing it into three beam paths for trapping, state-selection and state-detection. Each beam path contains a double-pass acousto-optic modulator (AOM) for intensity and frequency control. Trapping light is coupled into an optical fiber, the output of which is incident on the grating through an aperture in the top of the cavity after expansion and collimation. State selection and detection light is coupled into a common additional fiber. From the fiber the light is linearly polarized with respect to the magnetic bias field and split with a non-polarizing beam splitter (NPBS). This allows a portion of the light to be monitored on a reference photodiode while the rest of the light is aligned through the side holes in the cavity body and retro-reflected by a mirror back onto a second signal photodiode. This allows normalization of the absorption signal for fluctuations in laser intensity during state-detection.

### b) Experimental Cycle

The experimental cycle is presented in Figure 1. The traditional Ramsey cycle is preceded by a trapping and cooling phase by optical molasses of duration $t_{load}$ where the intensity is linearly ramped down from $5 I_{sat}$ to $0.1 I_{sat}$[32] while simultaneously scanning the detuning from $-12\ MHz$ to $-30 MHz$. After this step $\approx 1 \times 10^6$ atoms are in the states $|F = 2, m_F = \pm 2, \pm 1, 0\rangle$ with temperature $\approx 10\ \mu$K. A state selection phase of duration $t_{prep}$ utilizing light of intensity $0.02 I_{sat}$ that is linearly polarized with respect to the C-field is then used to prepare the atoms in the clock state. This light is incident on the atoms through the cavity side-hole and is tuned to the $|F = 2\rangle \to |F' = 2\rangle$ with repump light driving the $|F = 1\rangle \to |F' = 2\rangle$ transition. Due to selection rules this pumps more than 95% of the atoms into the $|F = 2, m_F = 0\rangle$ level [33], increasing the signal contrast.

The atoms are then driven with two $\pi/2$ microwave pulses of duration $t_m = 167\ \mu$s separated by a free evolution (or Ramsey) time $T_R$. The population $P_2$ in level $|F = 2\rangle$ is estimated by measuring the absorption of a side beam tuned on $|F = 2\rangle \to |F' = 3\rangle$. This value is normalized by the absorption of the same beam with $\approx 5\%$ $|F = 1\rangle \to |F' = 2\rangle$ sideband. This second pulse recycles atoms initially shelved in the $F = 1$ ground-state back into the cycling transition and allows normalization for variation of the number of atoms present at each shot. The C-field is optimized for atom cooling and is kept constant during the clock cycle to avoid eddy currents.

## IV. RESULTS

### a) Rabi oscillations

Rabi oscillations (see Figure 9) are measured by setting $T_R = 0$ ms. To sweep the microwave pulse area, the microwave time is kept constant and the microwave power is swept. The opposite approach, constant power and sweeping microwave time is disregarded to avoid the influence of time dependent relaxation processes and atom losses. The first minimum of the Rabi oscillations is used to calibrate the input microwave power to generate $\pi/2$ pulses.

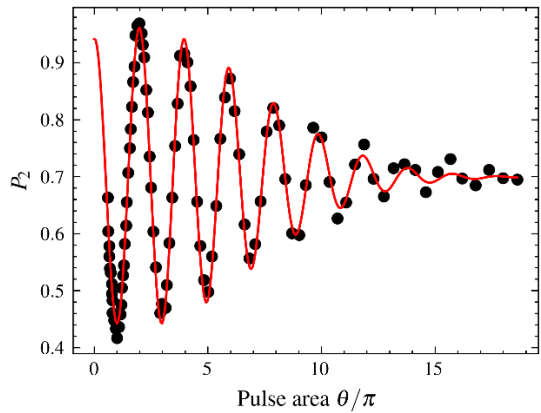

*Figure 9 - Rabi oscillations. The pulse area axis is rescaled assuming the first minimum is $\pi$.*

The decay of the Rabi oscillation is due to the residual magnetic field inhomogeneities inside of the atomic volume[34]. The number of visible oscillations, 9, is a sign of excellent $H_z$ homogeneity which is comparable to other compact cold atom clocks that show 14 or more oscillations[35], whereas high-performance vapor-cell clocks such as a $^{87}$Rb Ramsey maser show a number of oscillations as low as 3[36]. In the case of vapor-cell clocks, the atoms cover a larger volume inside the cavity, and the field homogeneity is dominated by the TE$_{011}$-like distribution. In contrast, the oscillation decay presented in Figure 9 is compatible with a pulse area $\theta$ with a gaussian distribution. Indeed, if one assumes a gaussian distribution with standard deviation $\sigma$, relative to the pulse area mean $\theta$, one can show that the form of the textbook $\cos^2(\theta/2)$ Rabi oscillations is modified as follows:

$$P_2(\theta) = \frac{A}{2}\left(1 + e^{-\sigma^2\theta^2/2}\cos(\theta)\right) + C, \quad (5)$$

with A and C two fit parameters that account for the absorption measurement. With this model, from a fit to Figure 9, we measure $\sigma = (6.3 \pm 0.1)\%$. The distribution of the field felt by the atoms is given by the convolution of the distribution of the atoms – which is assumed gaussian – and the distribution of the $TE_{011}$-like field. The smaller the volume spanned by the atoms, the more uniform the field distribution.

### b) Zeeman spectrum

The Zeeman spectrum is presented on Figure 10. It is intentionally acquired with repump light in the state selection beam extinguished, shelving atoms in the $|F = 1\rangle$ ground-state. The Ramsey cycle is set with $T_R = 0$ ms and with optimal microwave power deduced from the Rabi oscillations. Note that the amplitude of the $\pi$ transitions (i.e. transitions with $m_F = m'_F$) peaks at $\pm 1.1$ MHz are slightly asymmetric which is attributed to a small imbalance in the $\sigma^\pm$-polarized beams used for trapping and cooling. The Zeeman spectrum shows peaks instead of dips because in the absence of repump light, the majority of the atoms populate the sublevels $|F = 1, m_F = \pm 1, 0\rangle$ before the microwave cycle.

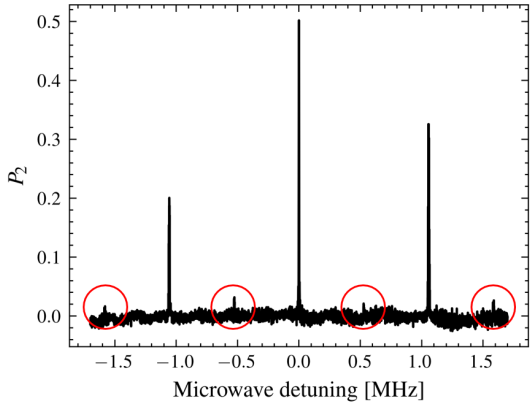

Figure 10 – Zeeman spectrum without state selection. The peaks circled in red are the $\sigma^\pm$ transitions.

Note that the $\sigma^\pm$ transitions (i.e. transitions with $m'_F = m_F \pm 1$) peaks are only slightly above the noise level which indicates excellent field homogeneity and alignment with C-field. The Zeeman spectrum allows to measure a field orientation factor $\xi$ of 97% which is a measure of field uniformity which was defined in [29], namely:

$$\xi = \frac{\int dv\, S_\pi}{\int dv(S_\pi + S_\sigma)} = \frac{\int_V d^3\vec{r} \cdot H_z^2}{\int_V d^3\vec{r} \cdot \vec{H}^2}, \quad (6)$$

with $S_\pi$ and $S_\sigma$ the Zeeman spectrum $\pi$ and $\sigma$ components only, respectively. $V$ is the volume spanned by the interrogated atoms.

### c) Ramsey fringes and expected clock performances

Ramsey fringes were obtained for Ramsey times up to 20 ms, limited in the current setup by the atoms falling under gravity and leaving the transverse optical beam after this time[37]. Figure 11 shows the optimal[37] SNR Ramsey fringes with $T_R = 10$ ms. The FWHM of the central fringe is 49 Hz and is compatible with eq. (1). The contrast is 58% due to the state selection. Without state selection the maximum contrast is 33%. The noise on the fringes is primarily attributed to magnetic field noise.

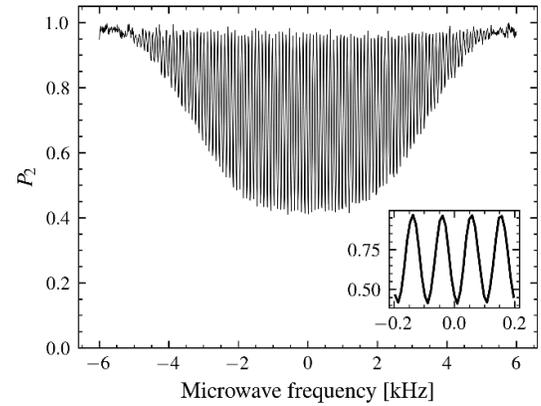

Figure 11 – Ramsey fringes for $T_R = 10$ ms and $t_m = 167\mu s$. The main figure and the inset share the same axes.

The nominal cycle time sequence is shown on Table II.
*Table II - Nominal time sequence.*

| Pulse | Duration |
|---|---|
| $T_{load}$ | 516 ms |
| $t_{prep}$ | 1 ms |
| $t_m$ | 167 μs |
| $T_R$ | 10 ms |
| $t_{det}$ | 1.4 ms |
| $T_C$ | 528.7 ms |

At one second, the Allan deviation of the future clock should be $\sigma(\tau = 1s) = 4.4 \times 10^{-11}$. The SNR of eq. (2) is estimated from a sine fit to the central fringe of Figure 11[38]. A preliminary estimation shows that the Dick effect contribution[39] should be at the level of $\sigma_{Dick}(\tau = 1s) = 1.2 \times 10^{-12}$. More detailed short-term analysis of increased performances clock is presented in a more advanced version of the setup[37]. Finally, we estimate the cavity-pulling (see eq. (4)) to be of the order of $2.1 \times 10^{-15}$ for a reasonable relative cavity temperature fluctuation of $10^{-3}$ which is typical at $10^5$ s.

## V. CONCLUSION

We demonstrated the successful design, simulation and implementation of a microwave cavity for a novel compact cold atom DR clock based on intra-cavity atom cooling using a diffractive optical element. An efficient simulation workflow allows us to design the cavity with a small mismatch of the resonance frequency which is easily compensated via small adaptation of the cavity height. Rabi oscillations and Zeeman spectrum show proof of good field uniformity and homogeneity, respectively.

Short-term performances of the future proof-of-concept clock are estimated to be close than that of miniature vapor-cell clocks[9,40], with a Dick effect limitation one order of magnitude lower. Accordant stability measurement has been demonstrated in an early stage clock setup with optimized signal[37]. In order to improve on the short-term stability of future setups, the Ramsey time could be increased by either putting the cavity upside down and interrogating the atoms further down or operating the clock in a mini-fountain fashion with Ramsey times up to 50 ms and 100 ms[16], respectively. This will also allow to interrogate the atoms in regions where the field is more uniform, as suggested by Figure 5. To improve the Dick effect limit, $T_C$ must be reduced which can be achieved by implementing atom recapturing at the end of the cycle[41,42].

## ACKNOWLEDGMENTS

E. Batori, C. Affolderbach and G. Mileti thank V. Dolgovskiy for his contributions in the early design phase, P. Scherler for technical assistance and G. Bourban for his continued support. This work was financially supported by the Swiss Space Center (Swiss Confederation) and the European Space Agency (ESA, contract 4000131046). The views expressed herein can in no way be taken to reflect the official opinion of the European Space Agency. A.B. was supported by a Ph.D. studentship from the Defence Science and Technology Laboratory (Dstl). We gratefully acknowledge support from the International Network for Microfabrication of Atomic Quantum Sensors (EPSRC EP/W026929/1).

## DATA AVAILABILITY

The data that supports the findings of this study are openly available in b2share at doi.org/10.23728/b2share.cb87f03a908343aeade769b3d054355e, reference number cb87f03a908343aeade769b3d054355e.